\documentclass{aa}  

\usepackage{color}
\usepackage[colorlinks, allcolors=blue]{hyperref}
\usepackage{graphicx}
\usepackage{xcolor}
\usepackage{amsmath}
\usepackage{txfonts}
\begin{document}

   \title{Constraining the magnetic vector in the quiet solar photosphere and the impact of instrumental degradation}

   \author{R. J. Campbell
        \inst{1}
          \and
          S. Shelyag\inst{2}
          \and
          C. Quintero Noda\inst{3,4}
        \and
        M. Mathioudakis
          \inst{1}
        \and
        P. H. Keys
          \inst{1}
        \and
        A. Reid\inst{1}
          }

   \institute{Astrophysics Research Centre (ARC), Queen's University of Belfast,
              Northern Ireland, BT7 1NN, UK\\
              \email{rcampbell55@qub.ac.uk}
         \and
             School of Information Technology, Deakin University Geelong, Australia
        \and
        Instituto de Astrof\'isica de Canarias, V\'ia L\'actea s/n, E-38205 La Laguna, Tenerife, Spain\vspace{-0.3cm}\\
        \and
        Departamento de Astrof\'isica, Univ. de La Laguna, La Laguna, Tenerife, E-38205, Spain
             }

   \date{Received May 28, 2021; accepted July 3, 2021}

 
  \abstract
   {With the advent of next generation high resolution telescopes, our understanding of how the magnetic field is organized in the internetwork (IN) photosphere is likely to advance significantly.}
   {We aim to evaluate the extent to which we can retrieve accurate information about the magnetic vector in the IN photosphere using inversion techniques.}
   {We use snapshots produced from high resolution three-dimensional magnetohydrodynamic (MHD) simulations and employ the Stokes Inversions based on Response functions (SIR) code to produce synthetic observables in the same near infrared spectral window as observed by the GREGOR Infrared Spectrograph (GRIS), which contains the highly magnetically sensitive photospheric Fe I line pair at $15648.52$ $\AA$ and $15652.87$ $\AA$. We then use a parallelized wrapper to SIR to perform nearly $14$ million inversions of the synthetic spectra to test how well the `true' MHD atmospheric parameters can be constrained statistically. Finally, we degrade the synthetic Stokes vector spectrally and spatially to GREGOR resolutions and examine how this influences real observations, considering the impact of stray light, spatial resolution and signal-to-noise (S/N) in particular.}
   {We find that the depth-averaged parameters can be recovered by the inversions of the undegraded profiles, and by adding simple gradients to magnetic field strength, inclination and line of sight velocity we show that an improvement in the $\chi^2$ value is achieved. We also evaluate the extent to which we can constrain these parameters at various optical depths, with the kinematic and thermodynamic parameters sensitive deeper in the atmosphere than the magnetic parameters. We find the S/N and spatial resolution both play a significant role in determining how the degraded atmosphere appears. At the same time, we find that the magnetic and kinematic parameters are invariant upon inclusion of an unpolarized stray light. We compare our results to recent IN observations obtained by GREGOR. We studied a linear polarization feature which resembles those recently observed by GRIS in terms of appearing as `loop-like' structures and exhibiting very similar magnetic flux density. Thus, we demonstrate that realistic MHD simulations are capable of showing close agreement with real observations, and the symbiosis between them and observations continues to prove essential. We finally discuss the considerations that must be made for DKIST-era observations.}
   {}

   \keywords{Sun: photosphere --
                Sun: spectropolarimetry --
                infrared
               }

   \maketitle
%

\section{Introduction}
\begin{figure*}
\centering
\includegraphics[width=.9\textwidth]{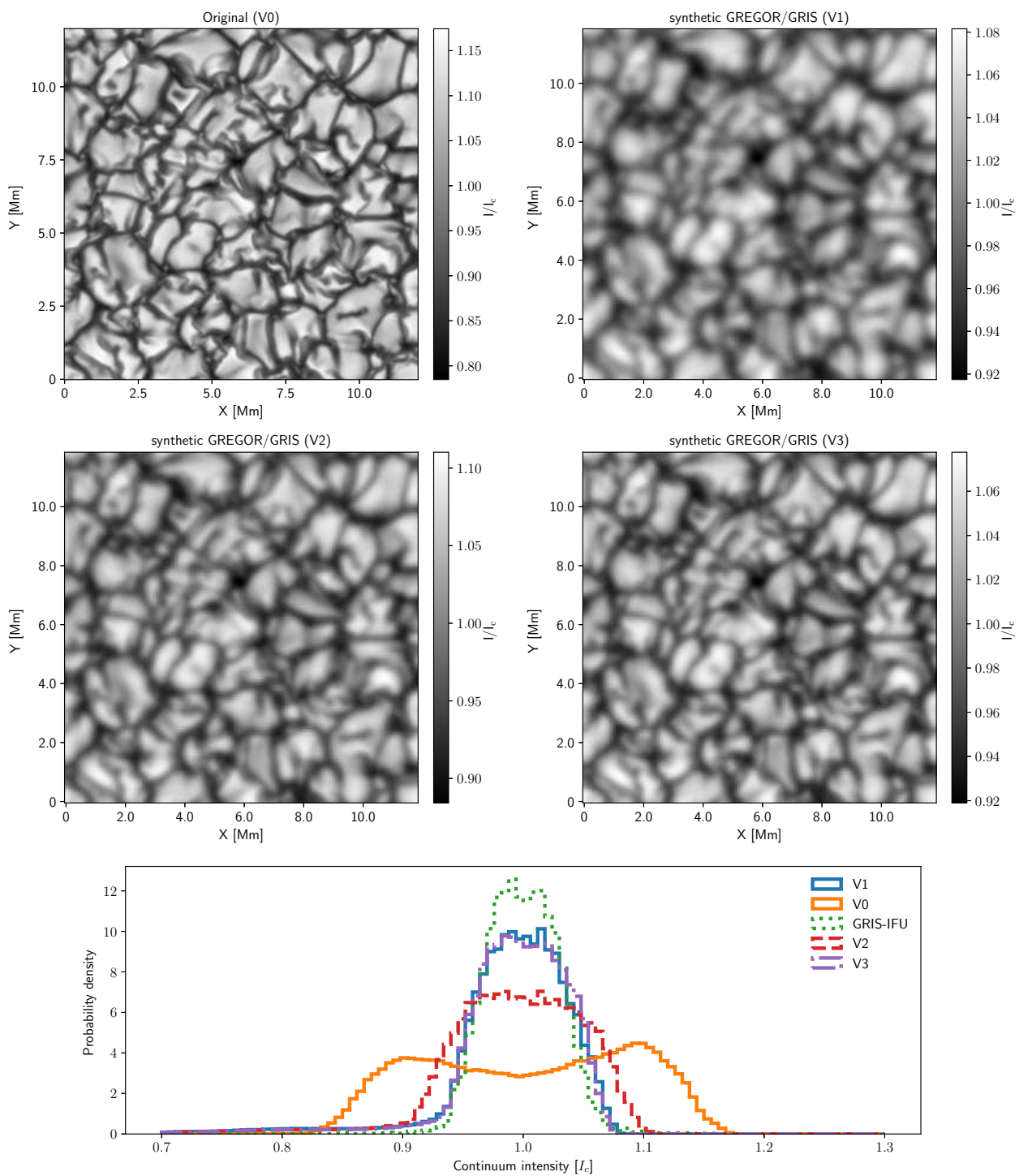}
\caption{Synthetic data produced from synthesis of MHD simulation outputs using SIR, shown at a continuum wavelength of 1565.47 nm. The \textit{top left} panel shows the original Stokes $I$ synthetic map (V0), while the \textit{top right} and \textit{middle left} panels show the spatially degraded map at an effective resolution of $380$ km (V1) and $215$ km (V2). The \textit{middle right} panel shows the same as V2 but further degraded by stray light (V3). The \textit{bottom} panel shows a histogram of the synthetic continuum intensities in each case, in addition to the distribution from real GREGOR/GRIS-IFU observations from $6$ May $2019$.}
          \label{fig:degraded_StokesI}%
\end{figure*}
As instrumentation improves, inversions of the increasingly large numbers of polarized Stokes profiles measured in observations of small-scale magnetic fields in the solar photosphere continues to be fraught with risks, limitations, and challenges (see \cite{rubio2019} for a full review). At spatial resolutions currently achievable, asymmetric and multi-lobed polarized Stokes vectors are common \citep{khomenko2003, kiess} in the internetwork (IN), indicating that we are not resolving the magnetic structures. To accurately reproduce these profiles, multiple components or gradients in optical depth are necessary, with both solutions providing very similar fits to the observations \citep{marian2016}. To retrieve accurate inclinations, one must measure both linear polarization and circular polarization above the noise level. As the amplitudes of these signals typically are very close to this noise level, the signal-to-noise ratio (S/N) has a huge impact on the distribution of inclinations retrieved \citep{andres2009}. Even more challenging is determining the azimuthal angle of the magnetic vector, which requires both Stokes $Q$ and $U$ above the noise level and utilization of disambiguation techniques. That is, to fully constrain the magnetism of the IN, full-vector spectropolarimetry is required.

Recent observations by \cite{Campbell} with the GREGOR infrared spectrograph integral field unit (GRIS-IFU) instrument \citep{GRIS} demonstrate that even with a very highly magnetically sensitive spectral line in the near infrared, observed with the largest solar telescope in Europe \citep{GREGOR1,GREGOR2}, it is possible to observe linear polarization in one frame, only for it to vanish in the next with a relatively short cadence. Therefore, with IFUs, a balance must be achieved between total integration time, field of view (FOV) size and cadence. These instrumental limitations are further compounded by Zeeman mixing induced by insufficient spatial and spectral resolution or, indeed, the evolution of the target during exposures. Observing the dynamics of small-scale magnetic fields has previously proved very challenging \citep{IMaX2013,Campbell,gosic2021}. It remains a key science goal to reveal and study the dynamics of magnetism in increasingly larger areas of the IN, as the emergence of flux in these regions can account for the transport of a large amount of energy to the solar surface \citep{nature2004}. Previous studies with very large FOVs have shown there are enormous regions of the IN apparently devoid of magnetic flux that can be revealed as harbouring a significant amount of transverse magnetic flux with sufficient S/N and spatial resolution \citep{Lites2008} or, indeed, with the Hanle effect \citep{stenflo1989}.

\begin{figure*}
    \centering
    \includegraphics[width=.9\textwidth
    ]{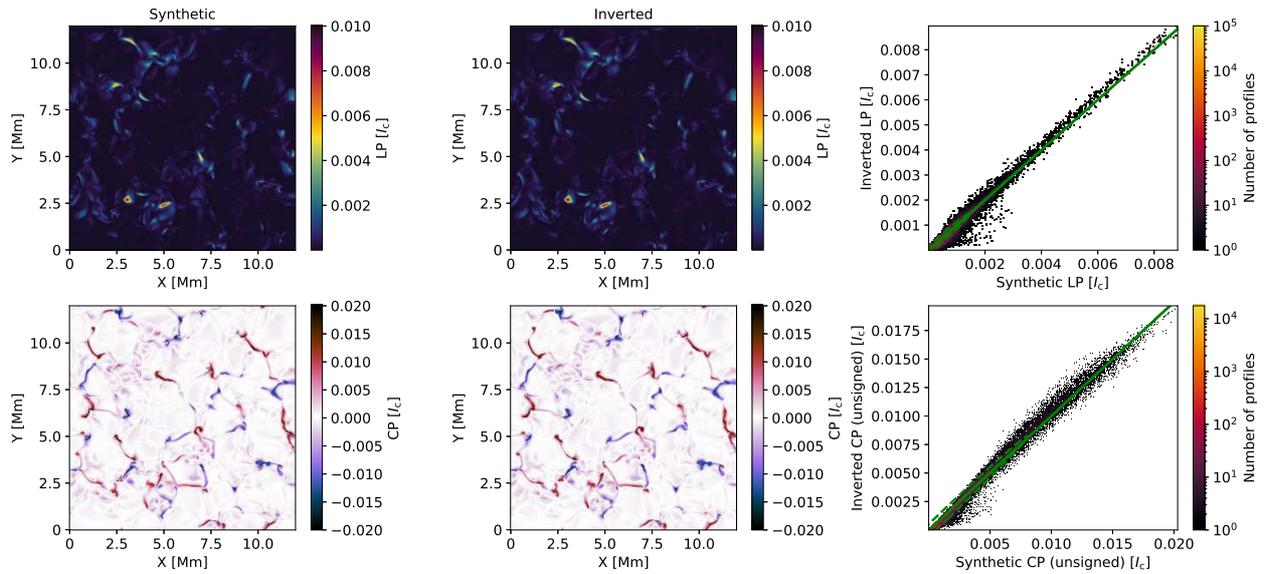}
    \caption{Comparison of the synthesized (\textit{left}) and inverted (\textit{middle}) maps of the LP (\textit{upper row}) and CP (\textit{lower row}) according to scheme G. For both LP and CP the synthesized and inverted parameters are plotted against each other with the colour scale on a $log_{10}$ scale (\textit{right}). LP and CP, defined in Eq. \ref{eqn:LP} and \ref{eqn:CP}, respectively, are computed by integrating across the $15648.52$ $\AA$ line. The \textit{solid, green} line indicates the exact linear relationship. The \textit{dashed, green} ellipse indicates the $99\%$ confidence level.} 
    \label{fig:syn_v_inv_obs}
\end{figure*}

One method for ensuring results derived from observations are well constrained is to compare them to semi-empirical models \citep{borrero2011} or realistic magnetohydrodynamic (MHD) simulations \citep{Vogler,khomenko2005,khomenko2007,lagg2016,Lites2017,milic2019,smitha2020}. With the advent of next generation high resolution telescopes, our understanding of how the magnetic field is organized in the IN photosphere is likely to experience a significant leap as observations reveal these regions of the solar atmosphere at unprecedented spatial resolution. In this study, we use a snapshot of a realistic MURaM simulation to evaluate the extent to which we can retrieve accurate information about the magnetic vector in the IN photosphere using inversion techniques. We synthesize the full Stokes vector by solving radiative transfer under the assumption of local thermodynamic equilibrium (i.e. the forward problem, see \cite{inversions} for a full review), only to then invert the synthetic spectra to establish the extent to which the original atmosphere differs according to the inversions. We then investigate how this model atmosphere may appear had it been observed at the spatial resolutions achievable by the GREGOR/GRIS-IFU. Similar processes have been employed to determine the effects of instrumentation on the analysis of wave phenomena \citep{keys2021}. We critically examine the results reported by \cite{Campbell} at this resolution, and also seek to help guide the approach taken by future observers considering this problem at higher resolution (e.g. with the Daniel K. Inouye Solar Telescope (DKIST); \cite{DKIST}).

\section{MHD cube and synthesis}

We use a snapshot of a realistic 3D MHD model simulation of solar magnetoconvection \citep{Vogler} for synthesizing the Stokes profiles. The MHD cube, described in detail by \cite{nelson2013}, has a spatial resolution of $25$ km in the plane of the solar surface and $14$ km in depth. The cube extends $12$ Mm $\times$ $12$ Mm in the $x$ and $y$ directions and $1.4$ Mm in the $z$ direction, or $480\times480\times100$ grid points, respectively. The simulation started with a plane parallel solar atmosphere and then purely hydrodynamical convection was allowed to develop until it reached a steady state. A `checkerboard', mixed polarity magnetic field (of magnetic field strength, $B$, of $\pm50$ G) was introduced. The snapshot was taken $30$ min after this point, allowing the magnetic field to disperse and the cancellation of opposite polarities resulted in a reduction of magnetic energy that resembles the conditions of the photospheric IN. The result is a cube with an abundance of mixed polarities, mostly concentrated in the intergranular lanes (IGLs).

\subsection{Transformation of the atmosphere to an optical depth scale and line synthesis}

We use the SIR code \citep[Stokes inversion based on Response Functions;][]{SIR} for synthesizing and inverting the full Stokes vector. We assume disk centre observations for all the studies we perform in this work (i.e. $\mu=1$, where $\mu=\cos(\theta)$ and $\theta$ is the heliocentric angle). We use the abundance values of the different atomic species given in \cite{Asplund2009}. We use the same atomic data as used by \cite{Campbell} and invert the same spectral lines. We do not degrade the Stokes profiles during the synthesis with any macroturbulence correction, and the microturbulence is null as in the original MURaM snapshot. 

We transform the 3D cube, originally in gas pressure and a geometrical height scale, to an atmosphere in optical depth using the subroutines found inside the SIR source code distribution. The transformation is a requirement of the SIR code for doing synthesis and inversions. To evaluate the optical depth at $5000$ $\AA$ from geometrical height and gas pressure we integrate the equation
\begin{equation}
   \mathrm{d}\tau_{5000\AA}=  -\rho k_{5000\AA}  \mathrm{d}z,   
\end{equation}
where $\rho$ is the density, $k_{5000\AA}$ is the continuum opacity at $5000~\AA$ and $z$ is the geometrical height. The $k_{5000\AA}$ is evaluated following the recipe of \cite{Wittmann1974}. The main difficulty of this computation is that we need to evaluate all the opacity contributions to $k_{5000\AA}$ and for that we need to know the density of each ion that has an important contribution to the continuum opacity (e.g., H$^-$, H, $\mathrm{He}^{-}$, He,  $\mathrm{H}_{2}^{-}$,  $\mathrm{H}_{2}^{+}$, and metals such as $\mathrm{Cl}^{-}$, C, Na, Mg). Those densities are computed by solving the Saha equation under the LTE approximation of SIR. However, the Saha equation needs, as input, the electron density to compute the atomic populations of the different ions, but the electron density depends on the ionization rate (i.e. on the results of the Saha equation for every ion). Thus, the problem needs to be solved iteratively to achieve a solution that is self-consistent. SIR does this using an algorithm proposed by \cite{Mihalas1967}, that is also described in \cite{Wittmann1974}.


\subsection{Degradation and Comparison with GREGOR observations}\label{section:degrade}
\begin{table}[b]
\caption{Percentage of synthetic linear (LP) and circular (CP) polarization profiles with maximums across the $15648.52\AA$ line above given thresholds, $\sigma_t$.}       
\label{table:polstats}      
\centering                          
\begin{tabular}{ c | c c c c }    
\hline
\hline
   Version & $\sigma_t$ [$I_c$]& $\%$LP & $\%$CP & $\%$LP + CP \\
 \hline                 

  V0 & $3.6\times10^{-3}$ & 19.5 & 77.5 & 18.7  \\
  V1 & $3.6\times10^{-3}$ & 0.91 & 42.8 & 0.76  \\
  & $2.4\times10^{-3}$ & 2.5 & 61.7 & 2.2  \\
  V2,3 & $3.6\times10^{-3}$ & 1.9 & 47.5 & 1.7  \\
   & $2.4\times10^{-3}$ & 4.4 & 63.8 & 4.0  \\
\hline
\end{tabular}\\ \vspace{.5cm}
\raggedright{\textbf{Note:} Values are shown for the various versions described in section \ref{section:degrade}. The percentages are calculated relative to the full FOV. These percentages are calculated before noise is added.}
\end{table}

\begin{figure*}
    \centering
    \includegraphics[width=.9\textwidth]{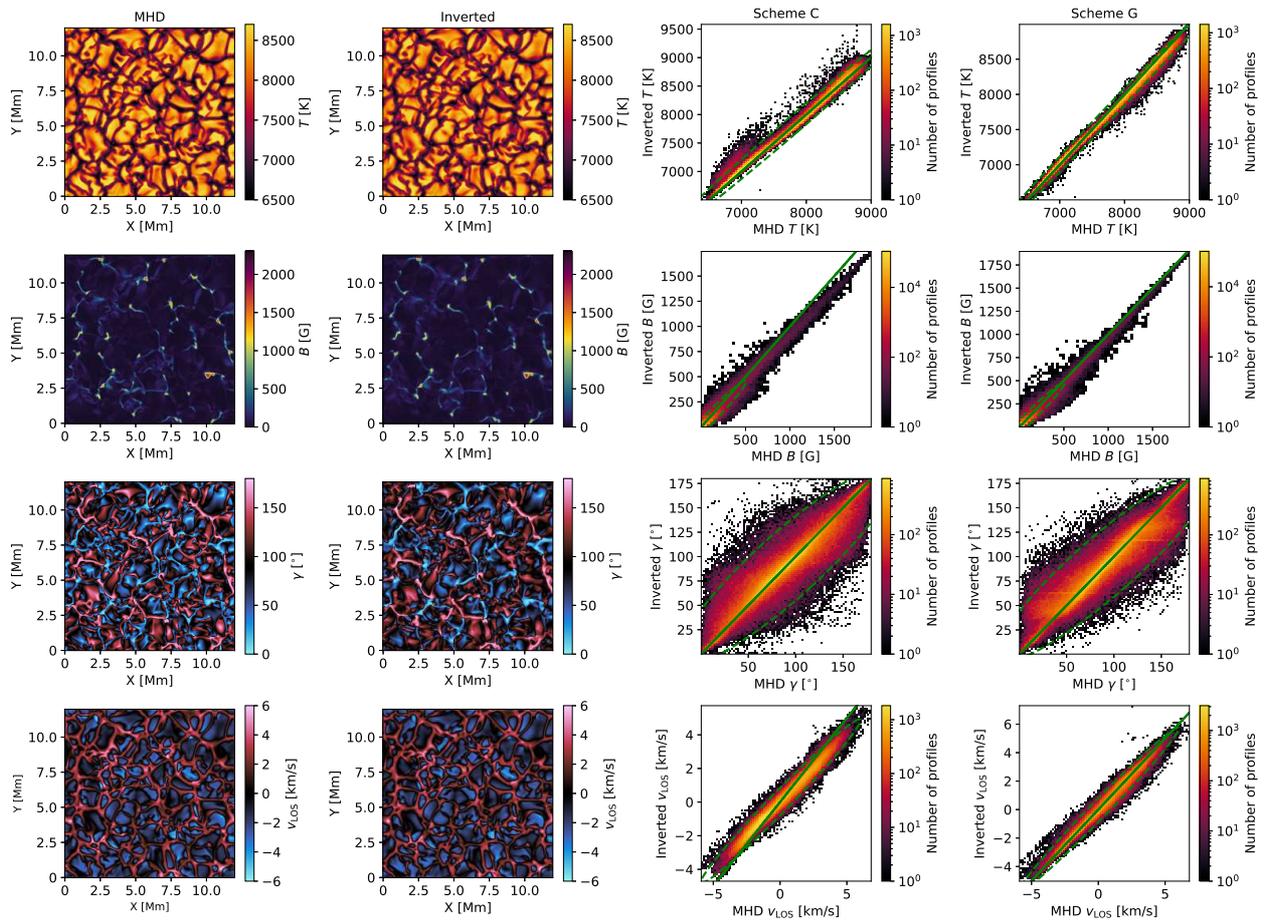}
    \caption{Comparison of the true MHD (\textit{first column}) and inverted (\textit{second column}) maps of the $T$ (\textit{top row}), $B$ (\textit{second row}), $\gamma$ (\textit{third row}) and $v_{\mathrm{LOS}}$ (\textit{bottom row}). The maps of inverted parameters are shown as determined by inversion scheme G (except for $\gamma$, which is shown as determined by scheme C). The colour table for $B$ is extended higher than the maximum value ($1915$ G) to accommodate those with colour vision deficiency. The MHD and inverted parameters, for both scheme C (\textit{third column}) and G (\textit{fourth column}) are plotted against each other with the colour scale on a $log_{10}$ scale. The \textit{solid, green} line indicates the exact linear relationship. The \textit{dashed, green} ellipse indicates the $99\%$ confidence level. $T$ is shown at $log\tau_{5000\mathrm{\AA}} = 0.5$, while the other parameters are shown at $log\tau_{5000\mathrm{\AA}} = -0.5$. }
    \label{fig:syn_v_inv_models}
\end{figure*}

In order to compare real observations to our synthetic profiles produced from the MHD cube, we first have to account for instrumental effects by degrading the spatial and spectral resolution of the synthetic profiles. We followed a similar procedure as \cite{milic2019}, but with the added benefit of having real observations with which to compare our synthetic spectra. First, we convolved the synthetic unpolarized and polarized spectra with the spatial point spread function (PSF) of the GREGOR telescope. To estimate this PSF, we first considered the root-mean-square continuum intensity contrast, defined as
\begin{equation}
    \delta I_{\mathrm{rms}} \approx \frac{\sqrt{\frac{1}{N}\sum_{x,y}\bigg(I(x,y) - \langle I_{x,y}\rangle \bigg)^2}}{\langle I_{x,y}\rangle},
\end{equation}
where $I(x,y)$ are continuum intensity values of individual pixels and $\langle I_{x,y}\rangle$ is the mean over all pixels, $N$. The peak $\delta I_{\mathrm{rms}}$ recorded by \cite{Campbell} with the GREGOR/GRIS-IFU was $3.1\%$ on $6$ May $2019$ with exceptional atmospheric seeing conditions. We used a least-squares procedure to iteratively convolve the synthetic cube with a two-dimensional Lorentzian kernel of variable scale parameter defining the width of the kernel, $\Gamma$, in Fourier space until the $\delta I_{\mathrm{rms}}$ of the convolved cube reached $3.1\%$. The $\Gamma$ value was determined as between $4-5$ pixels in the MHD cube, making the full width at half maximum (FWHM) about $0.53''$ or $380$ km on the solar surface. We then re-binned the degraded cube to a spatial sampling of $100$ km/pixel, which is approximately equivalent to the $0.135''$/pixel sampling of the GRIS-IFU in the $x$-direction. We refer to these degraded profiles as version 1 (V1). The theoretical diffraction limited resolution of GREGOR is $0.27''$ in this near infrared spectral window, but observations have not achieved this spatial resolution - $0.4''$ was reported by \cite{lagg2016}, but with a lower $\delta I_{\mathrm{rms}}$. We therefore followed the method described by \cite{milic2019} to consider the impact of stray light, which further degrades the $\delta I_{\mathrm{rms}}$. Here we are defining stray light as an additional unpolarized background added to the synthetic profiles, which the adaptive optics cannot correct. We added a very broad Gaussian to our spatial PSF, so that it becomes a Voigt profile. We model our spatial PSF as:
\begin{equation}
\begin{aligned}
    I^{\mathrm{stray}}(x,y) = (1-k) I^{\mathrm{spat}}(x,y) + k [I^{\mathrm{spat}}(x,y) * G(\sigma_{\mathrm{stray}})],
\end{aligned}
\end{equation}
where $I^{\mathrm{stray}}(x,y)$ is the spatially degraded and re-sampled cube of synthetic profiles, $I^{\mathrm{spat}}(x,y)$, further degraded by stray light, $k$ is the stray light fraction and $G(\sigma_{\mathrm{stray}})$ is a Gaussian with standard deviation, $\sigma_{\mathrm{stray}}$, of $2''$. Based on measurements by \cite{borrero2016} we used a $k$ value of 0.3. We then convolved the Voigt spatial PSF as before, keeping the $G(\sigma_{\mathrm{stray}})$ fixed and adjusting the $\Gamma$ value of the Lorentzian until we achieved a $\delta I_{\mathrm{rms}}$ of $3.1\%$ again. With this approach, the FWHM is measured as $0.3''$, much closer to the diffraction limited resolution of GREGOR. We refer to the synthetic profiles degraded (and re-sampled) at every wavelength with only the Lorentzian with FWHM of $0.3''$ as version 2 (V2), and the synthetic profiles degraded by the Voigt function, which additionally includes $G(\sigma_{\mathrm{stray}})$, as version 3 (V3). We also refer to the undegraded synthetic spectra as version 0 (V0). 

Although we produced the synthetic cube with the same dispersion as GREGOR/GRIS, we still need to consider the spectral resolution. The spectral PSF of GREGOR/GRIS has not been empirically measured, but estimations have been made with close agreement \citep{borrero2016, Campbell}. We convolved the synthetic cube in the wavelength domain with a Gaussian of FWHM $172$ m$\AA$. This also significantly reduced the amplitude of the polarized Stokes profiles. We applied this spectral degradation to V1, V2 and V3, and henceforth when we refer to any version (except V0) the inclusion of spectral degradation is implied. Finally, we added photon noise to each Stokes vector in wavelength space. At each spectral point, we added a value randomly derived from a normal distribution with a standard deviation of $\sigma_n = 8\times10^{-4}$ $I_{\mathrm{c}}$. This is the same as the noise level in Stokes $Q$ achieved by the GRIS-IFU in recent observations \citep{Campbell} and this will allow us to compare their results with the MHD simulations. We also separately considered a degraded cube with $\sigma_n = 3\times10^{-4}$ $I_{\mathrm{c}}$, which is similar to the noise level achieved by Hinode/SP in `sit-and-stare' mode \citep{Lites2008} and slightly below the noise level achieved by GRIS \citep{lagg2016}. The latter $\sigma_n$ value is meaningful as it allowed \cite{Lites2008} to reveal hidden magnetism in the IN, and we have calculated this as the lowest noise level the GRIS-IFU could achieve without the cadence between frames becoming too large to be sensible. This value should also be achievable by the the IFU at DKIST. In summary, V0 is the undegraded synthetic profiles. V1 is the synthetic profiles degraded to a spatial resolution of $0.53''$, re-sampled  to $0.135''$/pixel in $x$ and $y$ and spectrally degraded to $172$ m$\AA$. V2 is the synthetic profiles degraded to a spatial resolution of $0.3''$, spatially re-sampled and spectrally degraded in the same manner. Finally, V3 is the same as V2 but with additional stray light degradation.

Figure \ref{fig:degraded_StokesI} demonstrates the impact of the spatial PSF and stray light on the synthetic intensities. Much of the spatial detail differentiating IGLs from granules is lost. The distribution of continuum normalized intensities at GREGOR resolution is consistent with the distribution recorded by the GRIS-IFU for both V1 and V3. At the same time, V2 shows how if one wants to accurately model the spatial PSF based on measurements of $\delta I_{\mathrm{rms}}$ in ground-based observations, the spatial resolution will be dramatically higher if the impact of stray light is accounted for. This is important as we assume the stray light is unpolarized, meaning that if we neglect it we will under-estimate the spatial resolution and degrade the polarization signals by too large a degree, significantly reducing their amplitude while smearing the polarization over greater areas. Before instrumental degradation, the synthesized cube has $77.5\%$ of profiles with Stokes $V$ signals and $19.5\%$ with either Stokes $Q$ or $U$ displaying a maximum value in the $15648.52$ $\AA$ line greater than $3.6\times10^{-3}$ $I_{\mathrm{c}}$ (i.e. the $4.5\sigma_n$ level of the GRIS-IFU observations). Additionally, $18.7\%$ of profiles have a maximum in either Stokes $Q$ or $U$ and Stokes $V$ simultaneously above this amplitude. These values are reported in Table \ref{table:polstats}, along with the values for the various degraded versions at amplitudes of $3.6\times10^{-3}$ $I_{\mathrm{c}}$ and $2.4\times10^{-3}$ $I_{\mathrm{c}}$ (the latter being the $3\sigma_n$ level of the GRIS-IFU observations). The statistics for V2 and V3 are reasonably similar to that reported by \cite{Campbell} in terms of linear polarization, however with the synthetic profiles we have much more circular polarization than is measured in observations.

\section{Inversions}
\subsection{Inversions before instrumental degradation}
\begin{table}[b]
\caption{Number of free parameters (nodes) used in given atmospheric variables in the first and second inversion schemes (C and G, respectively).}       
\label{table:nodes}      
\centering                          
\begin{tabular}{ c | c c  c c c c c c }    
\hline
\hline
   &$T$ & $v_{mic}$ & $v_{LOS}$ & $v_{mac}$ & $B$ & $\gamma$ & $\phi$ & $\alpha$\\
 \hline                 

  Constant (C) & 4 & 0 & 1 & 0 & 1 & 1 & 1 & 0\\
  Gradients (G) & 4 & 0 & 2 & 0 & 2 & 2 & 1 & 0\\
\hline
\end{tabular}
\end{table}


     
    
      
      
      
\begin{figure*}
    \centering
    \includegraphics[width=.9\textwidth]{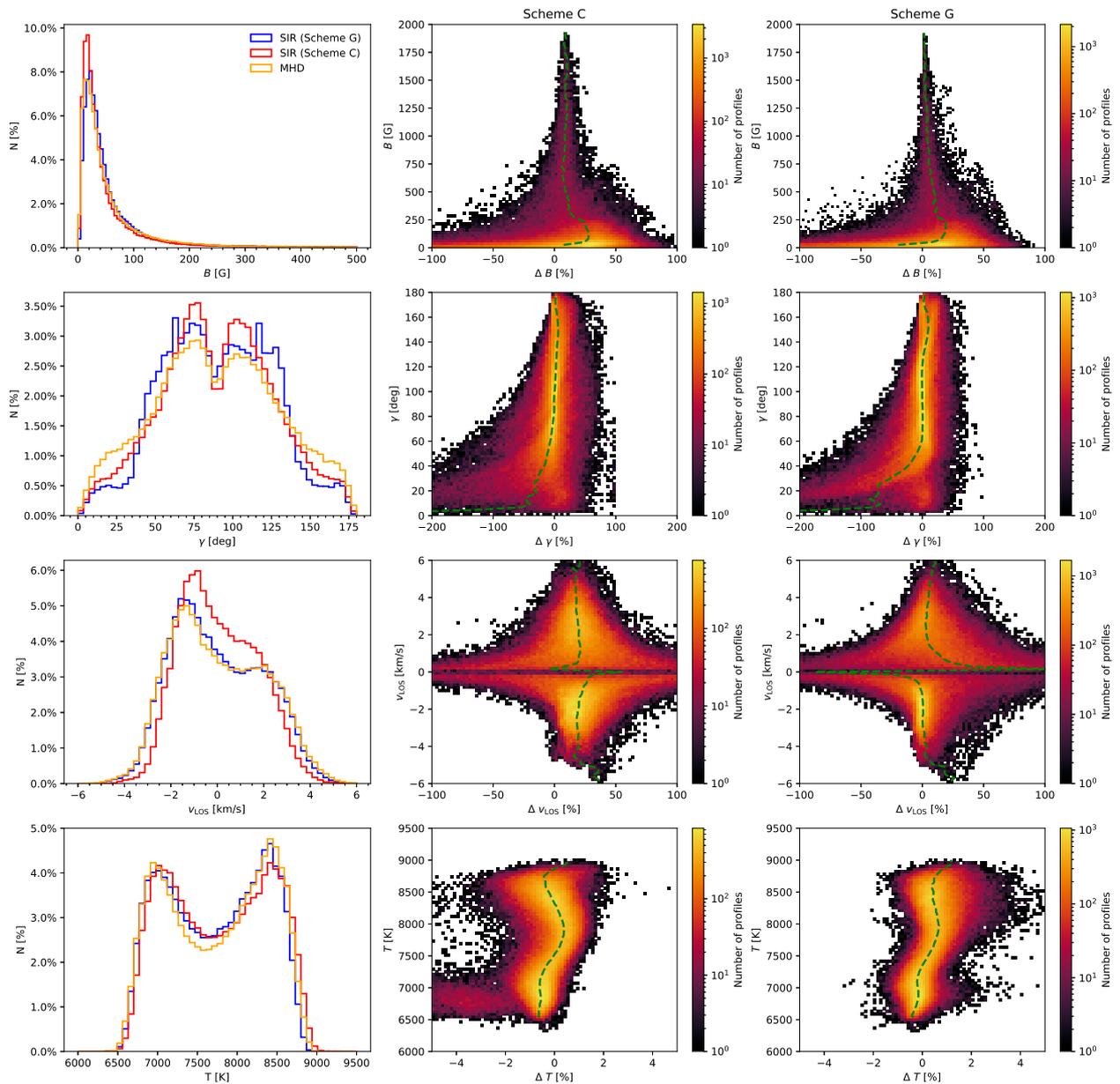}

    \caption{Distributions of values (\textit{left)} and histograms between the `true' MHD values and the differences between `true' MHD and inverted values (\textit{middle and right}, for scheme C and G, respectively) in $B$, $\gamma$, $v_{\mathrm{LOS}}$ and $T$ (\textit{upper to lower row}, respectively). Values are retrieved at $log\tau_{5000\mathrm{\AA}} = -0.5$, except for $T$ which is retrieved at $log\tau_{5000\mathrm{\AA}} = 0.5$. In the distributions, values from the MHD cube (\textit{orange lines}) and returned from the inversions under scheme C (\textit{red lines}) and G (\textit{blue lines}) are shown. The histograms of $B$, $\gamma$, $\Delta B$ and $\Delta \gamma$ are shown for pixels with maximum amplitude in at least one polarized Stokes vector $>2.4\times10^{-3}$ $I_{\mathrm{c}}$. The median error as a function of each parameter is shown by the \textit{green, dashed} lines. The percentages in all histograms are weighted with respect to the total number of pixels, N, i.e. including those with no polarization. The colour scale is shown on a $log_{10}$ scale.}
    \label{fig:undegraded_stats}
\end{figure*}

In order to test how well we can infer the thermodynamic, kinematic and magnetic properties of the solar atmosphere we inverted the synthetic spectra with the least-squares SIR code \citep{SIR}. We adopt a similar Monte Carlo approach to \cite{Campbell} in that we repeat the inversion procedure $30$ times per pixel with randomized model parameters in the inclination of the magnetic field with respect to the observer’s line-of-sight (LOS), $\gamma$, the azimuth of the magnetic field in the plane perpendicular to the observer’s LOS, $\phi$, the modulus of the magnetic field, $B$, and the line-of-sight velocity, $v_{\mathrm{LOS}}$. We followed this strategy to reduce as much as possible the probability of converging at a solution that is located in a local, as opposed to global, minimum in the $\chi^2$ hyper-surface, and this also reduces the influence of the initial guess model on the final solution. For the remaining atmospheric parameters, namely temperature, $T$, electron pressure, gas density and gas pressure, we initialise the inversion code with an initial FAL-C quiet Sun model \citep{font2006}, which describes each of the parameters characterizing the stratification of the atmosphere in the log of the optical depth at $5000$ $\AA$, $log\tau_{5000\mathrm{\AA}}$. We assume the magnetic filling factor, $\alpha$, is $1$, because in MURaM each pixel has no unresolved magnetic components. In this way, we are performing the experiment under the assumption that the resolution of the MHD cube represents the absolute scale of the Sun's atmosphere. We use the parallelized wrapper to SIR developed and made available by \cite{Ricardo2021}. By inverting multiple lines, with varying strengths, opacities and effective Land\'e g-factors, we can probe the atmosphere more reliably than if we inverted only one line. Table \ref{table:nodes} summarises the nodes used in the first inversion scheme (scheme C), which is a simple Milne-Eddington-like inversion, allowing us to retrieve depth-averaged parameters from the atmosphere. We chose $4$ nodes in $T$. In principle, the exact number of nodes in $T$ must be high enough to match the realistic stratification of the quiet Sun but low enough to avoid oscillatory solutions being introduced, which we do not observe in any cases in our results. We do not allow the microturbulent and macroturbulent velocities, $v_{mic}$ and $v_{mac}$, which account for the presence of small-scale and large-scale turbulent motions, respectively, to vary as free parameters. In the second inversion scheme (scheme G), we increase the nodes in $B$, $\gamma$ and $v_{\mathrm{LOS}}$ to $2$, to evaluate both the extent to which we can retrieve accurate atmospheric parameters at more than one optical depth and the extent to which the inclusion of a simple linear gradient can improve the $\chi^2$ values achieved. Using response functions, presented by \cite{Campbell} and \cite{Carlos2021}, it is expected that the greatest response to $B$, $\gamma$ and $v_{\mathrm{LOS}}$ is to be found at optical depths between $log\tau_{5000\mathrm{\AA}} = -1.0$ and $-0.5$. Meanwhile, the maximum response to $T$ is expected to be found at $log\tau_{5000\mathrm{\AA}} = 0.5$.

Figure \ref{fig:syn_v_inv_obs} shows, for the synthesized and inverted profiles under scheme C, the wavelength-integrated, net linear (LP) and circular (CP) polarization, which we define as follows:
\begin{equation}\label{eqn:LP}
    \textrm{LP} = \frac{\int_{\lambda_b}^{\lambda_r} [Q^2(\lambda)+U^2(\lambda)]^{\frac{1}{2}}  d\lambda}{I_c \int_{\lambda_b}^{\lambda_r} d\lambda},
\end{equation}
    
\begin{equation}\label{eqn:CP}
    \textrm{CP} = \textrm{sgn}(V_{\textrm{b}}) \frac{\int_{\lambda_b}^{\lambda_r} |V(\lambda)|  d\lambda}{I_c \int_{\lambda_b}^{\lambda_r} d\lambda},
\end{equation}
where $\textrm{sgn}(V_{\textrm{b}})$ represents the sign of the blue Stokes $V$ lobe and $\lambda_r$ and $\lambda_b$ are the red and blue wavelength limits of integration, respectively. It is clear that there is strong correlation for the each of these parameters between the synthetic and inverted profiles. The LP and CP are highly sensitive to the shape and amplitude of the lobes in Stokes $Q$ and $U$ in the former case and Stokes $V$ in the latter case. As scheme C can only produce symmetrical polarized Stokes profiles, a small difference is therefore expected.

\begin{table*}
\caption{Mean, standard deviation and median values in the percentage differences between values of $\gamma$, $B$, $v_{\mathrm{LOS}}$, and $T$ in the MHD cube and those returned from scheme C and G inversions. }       
\label{table:undegraded_errors}      
\centering                          
\begin{tabular}{ c | c  c  c | c  c  c}
\hline\hline
  \multicolumn{1}{c}{}& \multicolumn{3}{|c|}{Scheme C} & \multicolumn{3}{c}{Scheme G}  \\
                
   & Mean & Standard deviation & Median & Mean & Standard deviation & Median \\ 
\hline 
    $\Delta\gamma$ [$\%$] & -8.0 & 130.5 & 19.9 & -28.3 & 146.6 & 4.5\\
    $\Delta B$ [$\%$] & -9.5 & 58.8 & -0.4 & -11.5 & 67.5 & -0.1\\
    $\Delta v_{\mathrm{LOS}}$ [$\%$] & 21.1 & 45.3 & 19.9 & 4.5 & 49.2 & 1.4\\
    $\Delta\ T$ [$\%$] & -0.3 & 0.9 & -0.2 & 0.3 & 0.6 & 0.2 \\

\hline
\end{tabular}
\end{table*}

\begin{figure}
    \centering
    \includegraphics[width=.9\columnwidth]{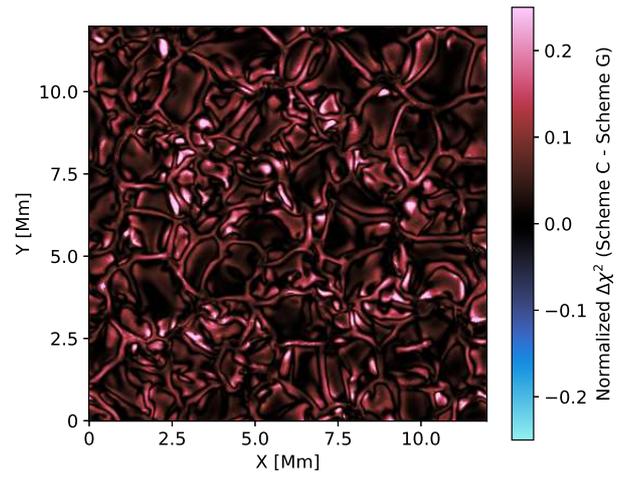}
    \caption{Map of the normalized difference between $\chi^2$ values produces by scheme C and scheme G inversions.}
    \label{fig:chi2}
\end{figure}

Figure \ref{fig:syn_v_inv_models} shows various atmospheric parameters for the synthesized and inverted profiles under scheme C and G. Values of $T$ and $\gamma$ compare very closely for both schemes, whereas values of $B$ and $v_{\mathrm{LOS}}$ appear slightly under-estimated by the inversion in scheme C. As is clear from the maps for all parameters, the topology of the parameters is well retrieved by the inversion. Fig. \ref{fig:undegraded_stats} further demonstrates this, showing the distributions of $B$, $\gamma$, and $v_{\mathrm{LOS}}$ at $log\tau_{5000\mathrm{\AA}} = -0.5$ for the input MHD atmosphere and those values returned by scheme C and G inversions. $T$ is also shown but at $log\tau_{5000\mathrm{\AA}} = 0.5$. Both C and G inversions reproduce the $B$ distribution well, but scheme C produces a $\gamma$ distribution that more closely resembles the MHD cube. In order to quantify how well constrained the parameters are, we calculate the percentage difference for each parameter. For temperature, we define the percentage difference, $\Delta T$, as
\begin{equation}
    \Delta T = \frac{T^{\mathrm{MHD}} - T^{\mathrm{SIR}}}{T^{\mathrm{MHD}}},
\end{equation}
where $T^{\mathrm{MHD}}$ is the value from the MHD cube transformed to the optical depth scale and $T^{\mathrm{SIR}}$ is the value returned by SIR under scheme C or G. The values of $T^{\mathrm{MHD}}$ and $T^{\mathrm{SIR}}$ are retrieved at the same optical depth. We similarly define $\Delta B$, $\Delta \gamma$, and $\Delta v_{\mathrm{LOS}}$. Positive values indicate that SIR has under-estimated the parameter, while negative values indicate an over-estimate. Fig. \ref{fig:undegraded_stats} also shows the percentage differences for both scheme C and G. It is important to emphasize that because the colour tables in Fig. \ref{fig:syn_v_inv_models} and Fig. \ref{fig:undegraded_stats} are on a $log_{10}$ scale, the vast majority of pixels are contained in the brightest bins in the histograms. Additionally, in Table \ref{table:undegraded_errors}, we calculate the mean, standard deviation and median of $\Delta T$, $\Delta B$, $\Delta \gamma$, and $\Delta v_{\mathrm{LOS}}$ across the populations in each case. When calculating $\Delta B$ and $\Delta \gamma$ we only consider pixels with a polarized Stokes vector with a maximum amplitude greater than $2.4\times10^{-3}$ $I_{\mathrm{c}}$ to restrict analysis to magnetic fields that would realistically be detectable. For $\Delta v_{\mathrm{LOS}}$, we consider only pixels with values $> 100$ ms$^{-1}$ to avoid very small values entering into the denominator, but $97\%$ of the FOV meets this criteria. In Fig. \ref{fig:undegraded_stats}, for each parameter, the median is overplotted as incrementally calculated across the populations in the same bins as the distributions are shown in the histograms. There is a small difference in percentage differences returned by the inversion schemes for $T$, but as this parameter had the same number of nodes in both schemes the improvement in scheme G is likely a by-product of the other parameters being better constrained. For $\gamma$, the values appear slightly better constrained by scheme C. It is further evident from the median values of $\Delta B$ and $\Delta v_{\mathrm{LOS}}$ in Table \ref{table:undegraded_errors} that scheme C under-estimates these parameters on average, but the reduction in the medians and the shifting of the distributions in Fig. \ref{fig:undegraded_stats} shows that scheme G resolves this issue, although there is still a spread of values. For $B$ in particular, stronger fields are better constrained than those in the weak field regime. As Fig. \ref{fig:chi2} illustrates, scheme G produces consistently lower  $\chi^2$ values than scheme C particularly in IGLs and in IGL-granule boundaries, where $v_{\mathrm{LOS}}$ and $B$ gradients are likely to be present.

\begin{figure}
    \centering
    \includegraphics[width=.9\columnwidth]{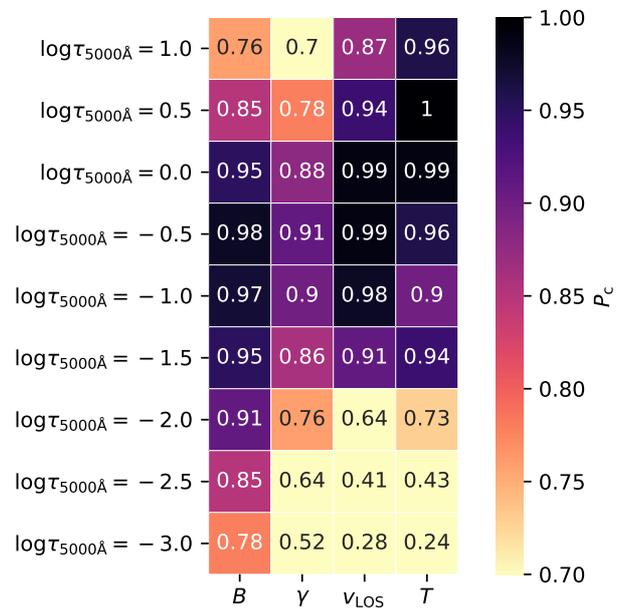}
    \caption{Pearson correlation coefficients for values of $B$, $\gamma$, $v_{\mathrm{LOS}}$ and $T$ between the MHD cube and scheme G estimates at a range of optical depths. The colour scale saturates at $0.7$. }
    \label{fig:pearson_MHDandG}
\end{figure}

\begin{figure*}
    \centering
    \includegraphics[width=.9\textwidth]{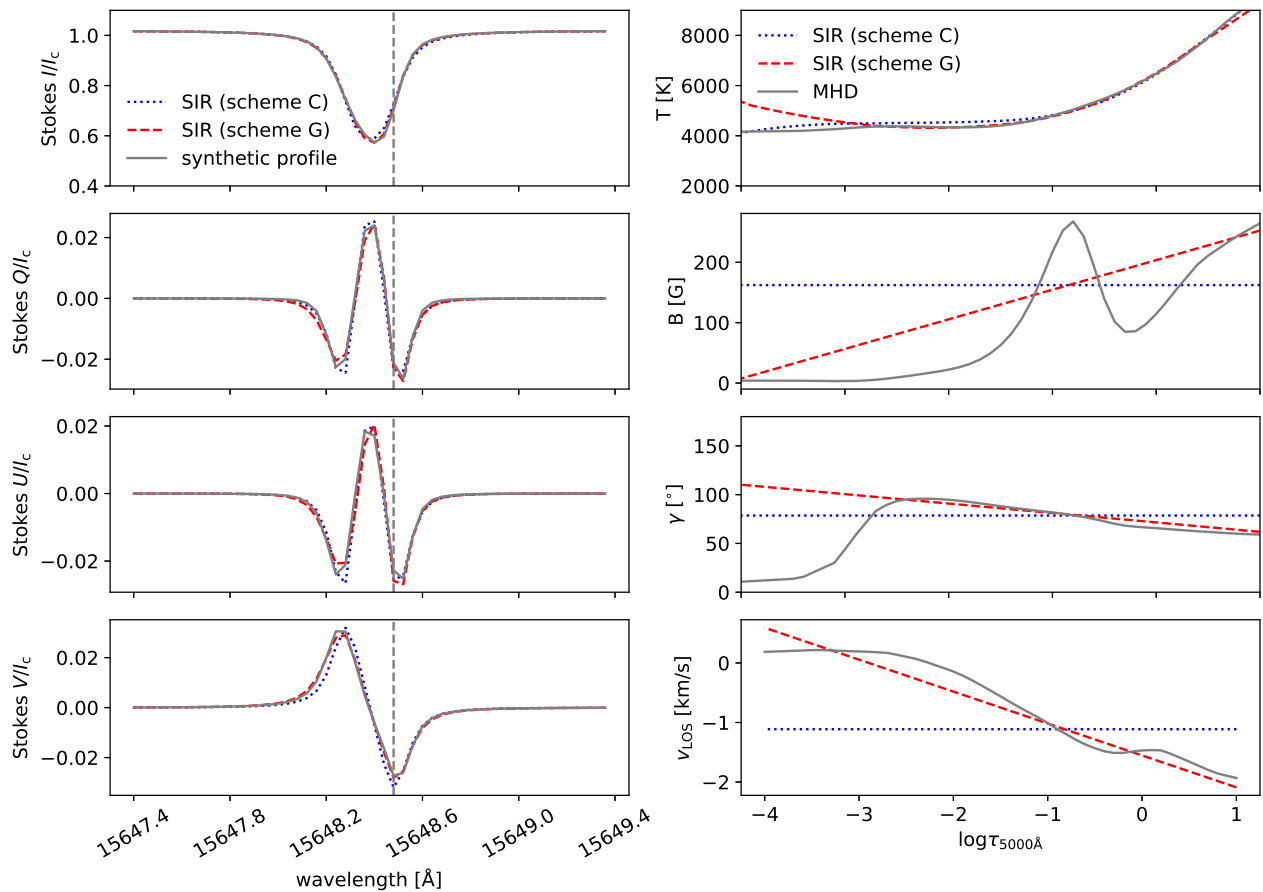}
    \caption{Sample synthetic profile with the inverted profiles and associated atmospheres. Stokes $I$, $Q$, $U$ and $V$ (\textit{grey, solid} lines) are shown for the $15648.52$ $\AA$ line (\textit{left panels}), normalised by the average continuum intensity, along with the inverted profiles from scheme C (\textit{blue, dotted} and G (\textit{red, dashed}) lines. The \textit{dashed, grey} line shows the rest wavelength. The retrieved atmospheric parameters (namely $T$, $B$, $\gamma$ and $v_{\mathrm{LOS}}$) are shown as a function of $log\tau_{5000\mathrm{\AA}}$ (\textit{right panels}). The stratification of the atmospheric parameters in the MHD model is also shown (\textit{grey, solid} line). 
    }
    \label{fig:example_profile_CandG}
\end{figure*}

We have established that the inclusion of simple gradients in the inversion schemes can improve the $\chi^2$ statistically, but in order to examine the extent to which we can retrieve accurate atmospheric parameters at more than one optical depth we calculated the Pearson correlations, $P_{\mathrm{c}}$, between the MHD values and the values returned by scheme G inversions. First, the values returned for scheme C were $1.00$, $0.99$, $0.98$, and $0.90$ for $T$, $v_{\mathrm{LOS}}$, $B$, and $\gamma$, respectively. Fig. \ref{fig:pearson_MHDandG} shows the $P_{\mathrm{c}}$ for scheme G at a range of optical depths. Of course, the results resemble very clearly the response functions, but we are explicitly testing the ability of the SIR code to retrieve the atmospheric parameters from realistic 3D MHD model outputs. The results confirm the potential for these lines to constrain the thermodynamic and kinematic parameters deep in the atmosphere, as also reported by \cite{milic2019}, with $T$ responsive to a broader range of optical depths than $v_{\mathrm{LOS}}$. In the deepest layers, the $P_{\mathrm{c}}$ is highest in $T$ as expected, since local thermodynamic equilibrium is firmly established in these layers \citep{jose}. $T$ also has two peaks in $P_c$, one very deep in the atmosphere, where Stokes $I$ is very responsive to $T$ in continuum wavelengths, and one higher, where wavelengths with spectral lines carry information about $T$. On the other hand,  the magnetic parameters do not appear to be as responsive to the deepest layers. For $B$, the $P_{\mathrm{c}}$ is highest between $log\tau_{5000\mathrm{\AA}} = -1.5$ and $-0.5$, but the correlation is also significant as high in the atmosphere as  $log\tau_{5000\mathrm{\AA}} = -2.0$ and as deep as $log\tau_{5000\mathrm{\AA}} = 0.0$. The $B$ values are less strongly correlated at $log\tau_{5000\mathrm{\AA}} = -2.0$ and below. For $\gamma$, the $P_{\mathrm{c}}$ is highest between $log\tau_{5000\mathrm{\AA}} = -1.0$ and $-0.5$, but the correlation is weaker than all other atmospheric parameters, highlighting the difficulty of constraining $\gamma$ in quiet Sun scenarios. It must be pointed out that these correlations could potentially be improved if nodes could be placed at specific optical depths based on an analysis of the response functions (as has been investigated by \cite{Navarro2004}) and associated errors. This functionality is not yet available with SIR but will be in the future (private communication with the developer).

A sample profile from a magnetic feature with significant linear polarization is shown in Fig. \ref{fig:example_profile_CandG}. In the MHD cube, this pixel had a very complex stratification in $B$, in addition to gradients in $\gamma$ and $v_{\mathrm{LOS}}$. The asymmetry in Stokes $V$ - which can be an observational indication of gradients in $v_{\mathrm{LOS}}$ - is present but very small. There are also small asymmetries in the $\sigma$-lobes of the linear polarization. Therefore, although scheme G does produce a better fit, and the approximation produced by SIR to the stratification of both $v_{\mathrm{LOS}}$ and $\gamma$ is reasonable, it is also clear that scheme C produced a very good fit without the inclusion of additional nodes. A much larger number of nodes would be required to model the stratification of $B$. The $T$ is very well constrained by both schemes. One can observe that scheme C and G naturally agree in their values of $B$, $\gamma$, and $v_{\mathrm{LOS}}$ at optical depths where the response functions peak, at optical depths between $log\tau_{5000\mathrm{\AA}} = -1.0$ and $-0.5$.

\subsection{Inversions at GREGOR resolutions}
At the resolution of the GREGOR telescope, we can assume that we are not able to resolve the small-scale magnetic structures we observe. For that reason, when inverting data observed at this resolution, we have to assume the resolution element (in the $x-y$ plane) contains more than one model. We then quantify the contribution of each model by the magnetic filling factor, $\alpha$. Additionally, we include $v_{mic}$ and $v_{mac}$ as free parameters to allow SIR to account for the spectral PSF. Table \ref{table:nodes1} describes the inversion that we employ for inverting the spatially and spectrally smeared synthetic cube. As in \cite{milic2019}, we observe in the degraded cube that the inferred velocities are lower and that downflows in IGLs are observed over a smaller area. In this section of the study, we focus our attention on the magnetic parameters - namely $B$ and $\gamma$. We are particularly concerned with the influence of noise on the retrieval of accurate magnetic parameters, especially $\gamma$. We cannot compare the original MHD values to the inverted parameters in specific pixels after degradation because the spatial and spectral PSF describe only their impact on observable quantities (i.e. on the Stokes vector) and it is highly likely that the impact on atmospheric parameters is non-linear. Therefore, instead, our approach is to first invert the degraded profiles and use the inverted atmosphere as a baseline. We compare the results of inversions of both V1 and V3, representing lower and upper estimates of the spatial resolution of the GRIS-IFU, and then attempt to establish the influence stray light has on the on retrieved parameters by comparing V2 with V3. We then add noise to the polarized Stokes vectors and invert again. We then apply the same principle component analysis (PCA) procedure as \cite{khomenko2003, marian2016, Campbell} and invert once more. We can then compare the atmospheric models of the noiseless, noisy and de-noised inversions and examine the impact of S/N on our results. In this way, we are also critically examining the effectiveness and correctness of the PCA procedure. 

\begin{table}
\caption{Number of free parameters (nodes) used in given atmospheric variables in inversion scheme S1; where a magnetic atmosphere is embedded with a field-free atmosphere.  }       
\label{table:nodes1}      
\centering                          
\begin{tabular}{ c  c  c }
\hline\hline
  Parameter & Model 1 & Model 2  \\   
\hline 
  T* & 4 & 4\\
 
  $v_{mic}$& 1 & 1 \\

  $v_{LOS}$& 1 & 1 \\
  
  $v_{mac}$*& 1 & 1\\
  
  $B$ & 0 & 1\\

  $\gamma$& 0 & 1\\

  $\phi$& 0 & 1 \\
  
  $\alpha$& 1 & 1\\
  
\hline
\end{tabular}\\ \vspace{.5cm}
\raggedright{\textbf{Note:} The asterisks (*) signify that a parameter is forced to be the same in both models.}
\end{table}
 

  
  


  
  

\begin{figure*}
    \centering
    \includegraphics[width=.3\textwidth]{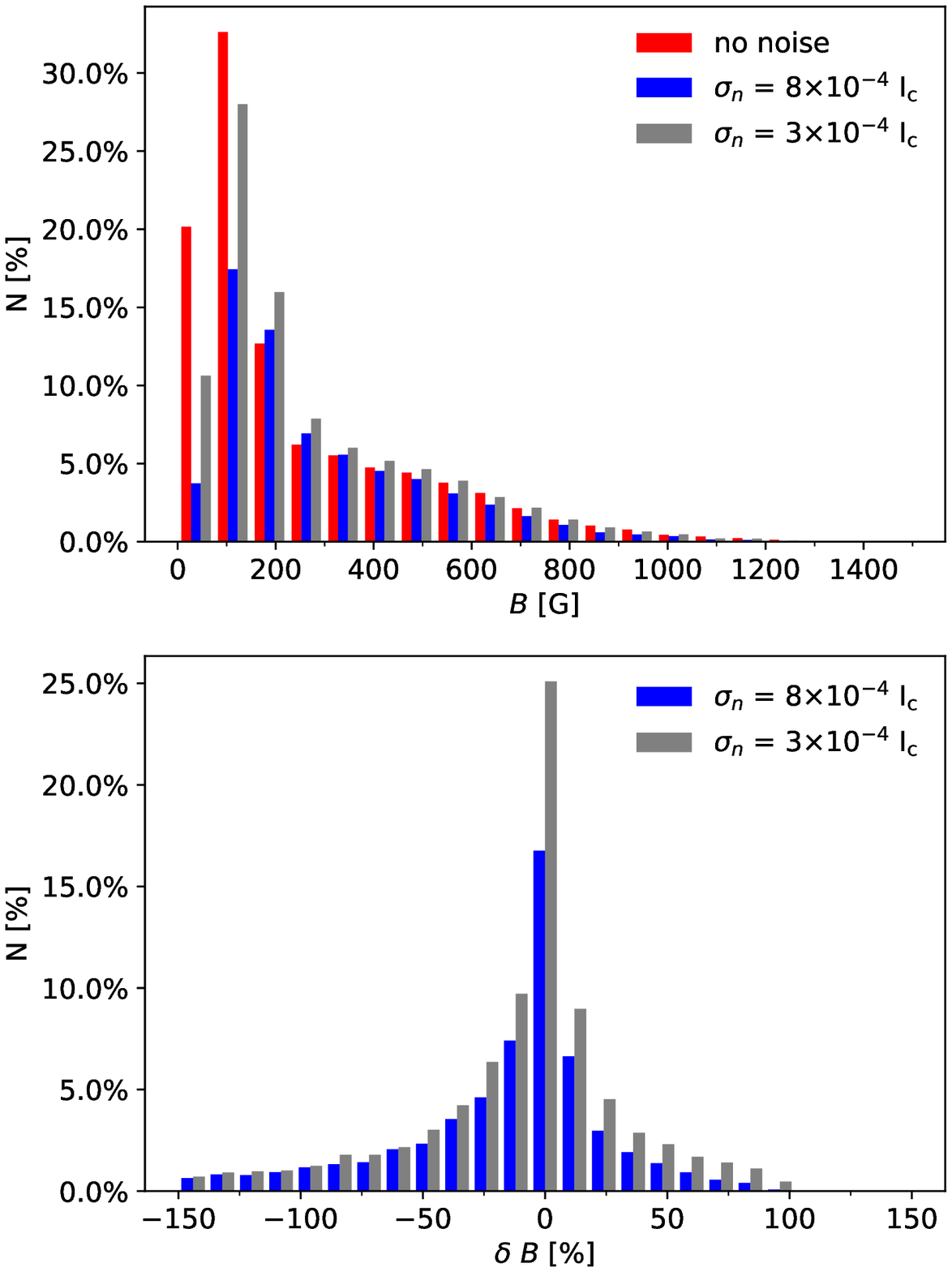}
    \includegraphics[width=.3\textwidth]{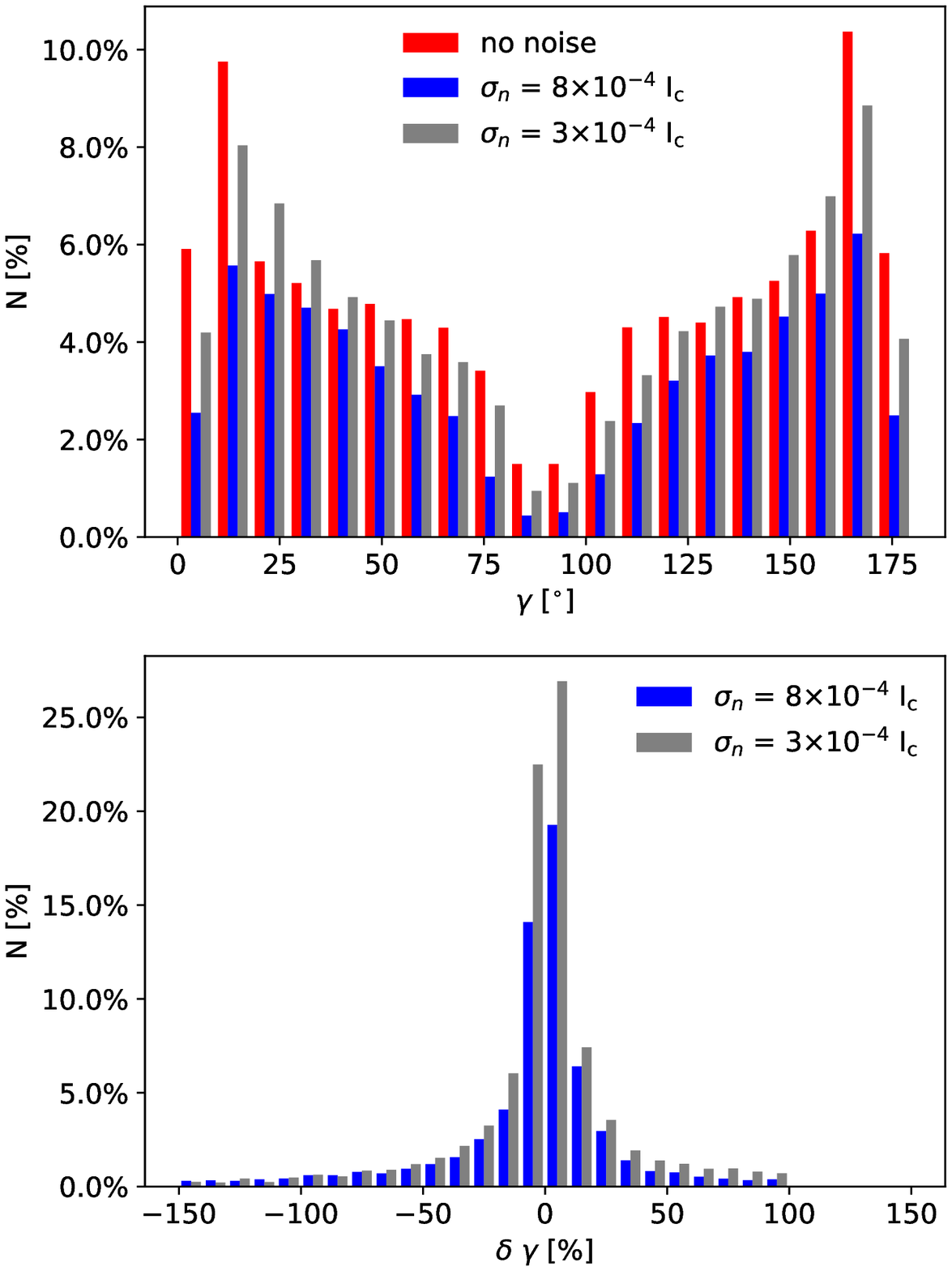}
    \includegraphics[width=.3\textwidth]{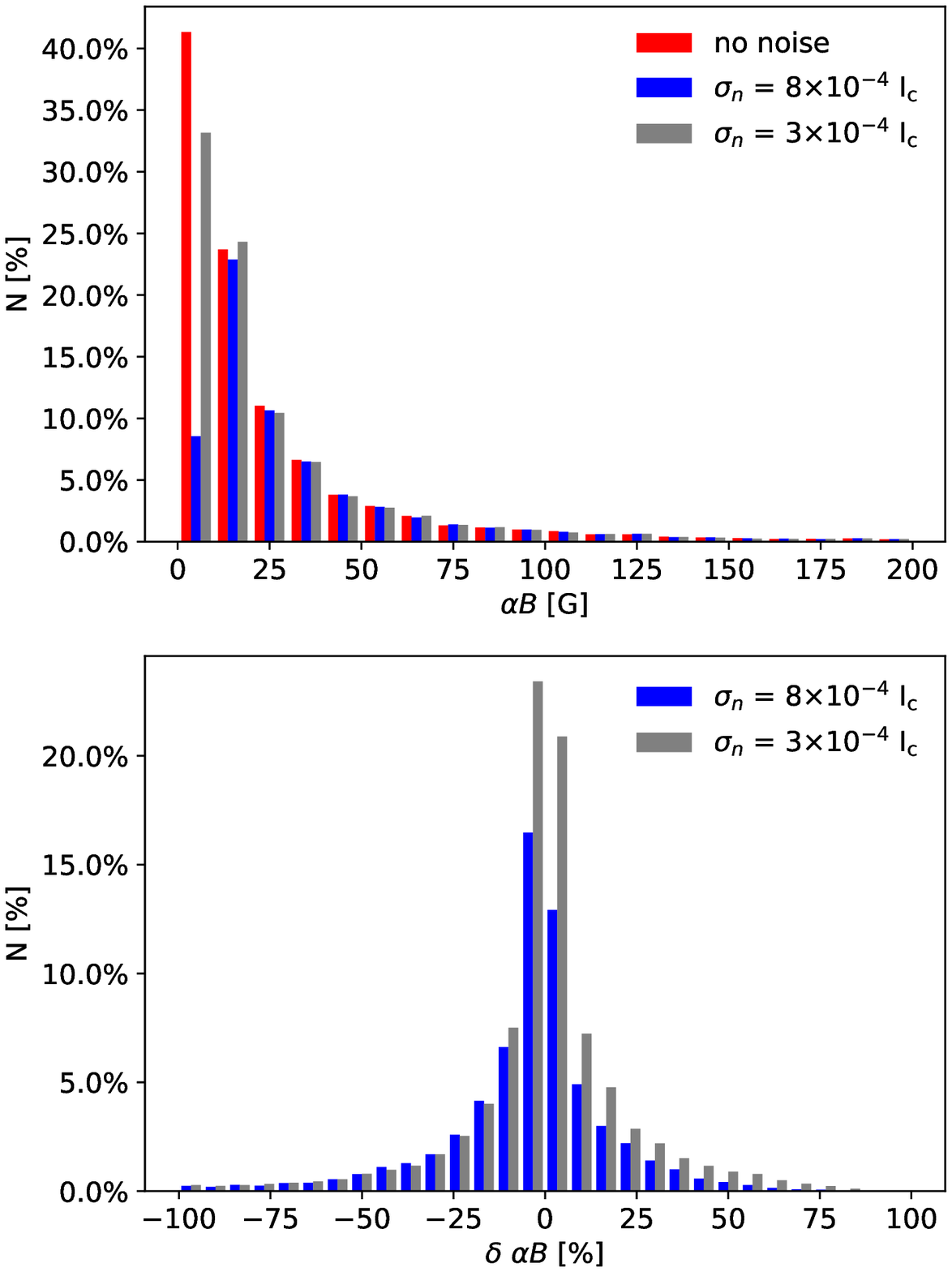}
    \caption{Distributions of B (\textit{left}), $\gamma$
 (\textit{middle}), and $\alpha B$ (\textit{right}) returned from the inversions under S1 at GREGOR resolutions (V3). The inversions were run without the addition of noise (\textit{red}), and then repeated with noise levels of $8\times10^{-4}$ $I_{\mathrm{c}}$ (\textit{blue}) and $3\times10^{-4}$ $I_{\mathrm{c}}$ added (\textit{grey}). The histograms of the inversions with noise are shown for pixels with maximum amplitude in at least one Stokes vector $>3\sigma_n$. The \textit{lower} row shows the difference between the values returned from inversions of the noiseless case and the values returned when noise is added. The percentages in all histograms are weighted with respect to the total number of pixels, i.e. including those with no polarization.}
    \label{fig:degraded_stats}
\end{figure*}

Figure \ref{fig:degraded_stats} shows the distributions of $B$, $\gamma$, and $\alpha B$ returned by the inversion under S1 for the noiseless and noisy V3 datacubes. On immediate inspection of the $B$ distribution, compared to Fig. \ref{fig:undegraded_stats}, we can observe the general impact of the smeared polarization profiles on inferred values; the strongest magnetic elements have been smeared over larger areas, resulting in a higher proportion of pixels with $B$ values in excess of $100$ G. The distribution of $B$ and $\alpha B$ is very similar to that reported by \cite{marian2016} and \cite{Campbell}. Regarding the $\gamma$ distribution, we know the vast majority of magnetic fields in the MHD cube have inclined (transverse) components, but as the amplitude of the relatively weak linear polarization signals are reduced by smearing, the degraded inverted atmosphere is much more vertical. In particular, the population of the highly inclined fields are impacted to a much larger extent than the intermediately inclined fields. Nevertheless, there remains significant proportion of fields with transverse components. 

Figure \ref{fig:degraded_stats} also shows the percentage differences between the $\gamma$, $B$, and $\alpha B$ values for inversions of the noiseless and noisy inversions at the two noise levels. Using the noiseless case as the baseline, we define the percentage differences between the $B$ values, $\delta B$ as,
\begin{equation}
    \delta B = \frac{B^{\mathrm{noiseless}} - B^{\mathrm{noise}}}{B^{\mathrm{noiseless}}}
\end{equation}
where $B^{\mathrm{noiseless}}$ is the value returned by SIR under S1 with no noise degradation and $B^{\mathrm{noise}}$ is the value when a given level of noise is added, quantified by $\sigma_n$. We similarly define $\delta\gamma$ and $\delta\alpha B$. It is clear that noise introduces an error into each of these parameters, and that with a lower noise level (higher S/N) we access polarization in a larger fraction of the FOV. With a higher S/N, a larger proportion of fields with strengths $<200$ G and magnetic flux densities $<25$ G are accessible. The distribution of $\delta B$ in particular is asymmetric, suggesting that the error induced by noise results in an over-estimation more often than an under-estimation. However, the distributions of $\delta\gamma$ and $\delta\alpha B$ are more symmetric. Table \ref{table:stds} shows the median of the absolute values in  $\delta B$, $\delta\gamma$, and $\delta\alpha B$ with both noise levels for V3 when only pixels with maximum amplitude in at least one Stokes vector $>3\sigma_n$ in the  $15648.52$ $\AA$ line are included, while Table \ref{table:stds2} shows the same for $4.5\sigma_n$. It is clear that the inclusion of noise results in deviations from the noiseless case. It is also clear that the median in $|\delta B|$, $|\delta\gamma|$, and $|\delta\alpha B|$ is slightly reduced with a larger noise threshold, but this improvement is small. In terms of $B$, while the median is less than $20\%$, the deviation can be as large as $50\%$ or higher, which is not a negligible deviation. However, the deviation in $\alpha B$ is smaller. Similarly, a deviation in $\gamma$ as large as $25-50\%$ is not uncommon, although the median is less than $10\%$. This validates the decision to characterize the field inclinations in broad terms (i.e. not in terms of precise values, but instead in terms of whether they appear highly vertical, highly inclined, or intermediate, as in \cite{Campbell}). At the same time, we found no evidence that PCA alters the values in any statistically significant way - the distributions for the inverted atmosphere after the application of PCA looked identical - showing that when a sensible number of eigenvectors is selected it can be safely applied to the Stokes vector without modification of the inferred atmospheres.

\begin{table}
\caption{Absolute median values in the percentage differences between $\gamma$, $B$ and $\alpha B$ returned from noiseless inversions and inversions with noise added at $\sigma_n = 8\times10^{-4}$ $I_{\mathrm{c}}$ and $\sigma_n = 3\times10^{-4}$ $I_{\mathrm{c}}$ to the spatially and spectrally degraded profiles (V3).}       
\label{table:stds}      
\centering                          
\begin{tabular}{ c  c  c }
\hline\hline
                
   & $\sigma_n = 8\times10^{-4}$ $I_{\mathrm{c}}$ & $\sigma_n = 3\times10^{-4}$ $I_{\mathrm{c}}$ \\ 
\hline 
    $|\delta\gamma|$ [$\%$] & 9.7 & 8.3\\
    $|\delta B|$ [$\%$] & 20.7 & 19.6 \\
    $|\delta\alpha B|$ [$\%$] & 8.3 & 7.1\\

\hline
\end{tabular}\\\vspace{.5cm}
\raggedright{\textbf{Note:} Only pixels with maximum amplitude in at least one Stokes vector $>3\sigma_n$ are included. }
\end{table}

\begin{table}
\caption{As in Table \ref{table:stds}, but only pixels with maximum amplitude in at least one Stokes vector $>4.5\sigma_n$ are included. }       
\label{table:stds2}      
\centering                          
\begin{tabular}{ c  c  c }
\hline\hline
                
   & $\sigma_n = 8\times10^{-4}$ $I_{\mathrm{c}}$ & $\sigma_n = 3\times10^{-4}$ $I_{\mathrm{c}}$ \\ 
\hline 
    $|\delta\gamma|$ [$\%$] & 8.0 & 7.5\\
    $|\delta B|$ [$\%$] & 15.0 & 17.8 \\
    $|\delta\alpha B|$ [$\%$] & 5.6 & 6.2\\

\hline
\end{tabular}
\end{table}

\begin{figure*}
    \centering
    \includegraphics[width=.9\textwidth]{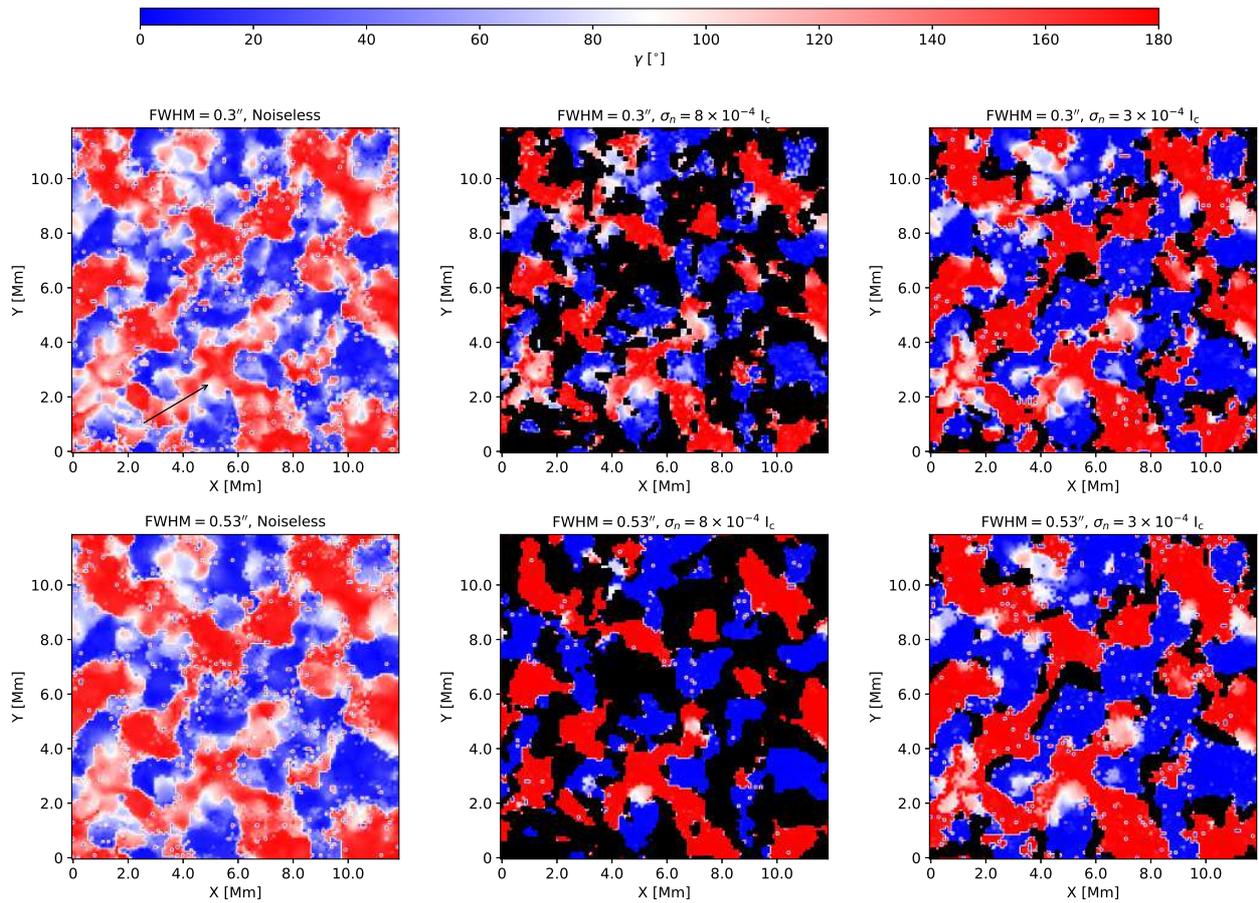}
    \caption{Maps of inclinations returned from S1 inversions for the spatially and spectrally degraded synthetic profiles, with no noise (\textit{left}) and two different noise levels ($8\times10^{-4}$ $I_{\mathrm{c}}$, \textit{middle}, and $3\times10^{-4}$ $I_{\mathrm{c}}$, \textit{right}). Only Stokes vectors with a maximum amplitude $>3\sigma_n$ are retained after PCA.  Black areas are representative of pixels which do not reach this threshold. The \textit{lower row} shows the maps for V1 (FWHM $= 0.53''$) while the \textit{upper row} shows the same for V3 (FWHM $= 0.3''$, with stray light included in the estimation and application of spatial PSF). The arrow in the \textit{upper left} panel indicates the spatial location of the sample profile shown in Fig. \ref{fig:Degraded_profile}.}
    \label{fig:degraded_inclinations}
\end{figure*}

\begin{figure*}
    \centering
    \includegraphics[width=.9\textwidth]{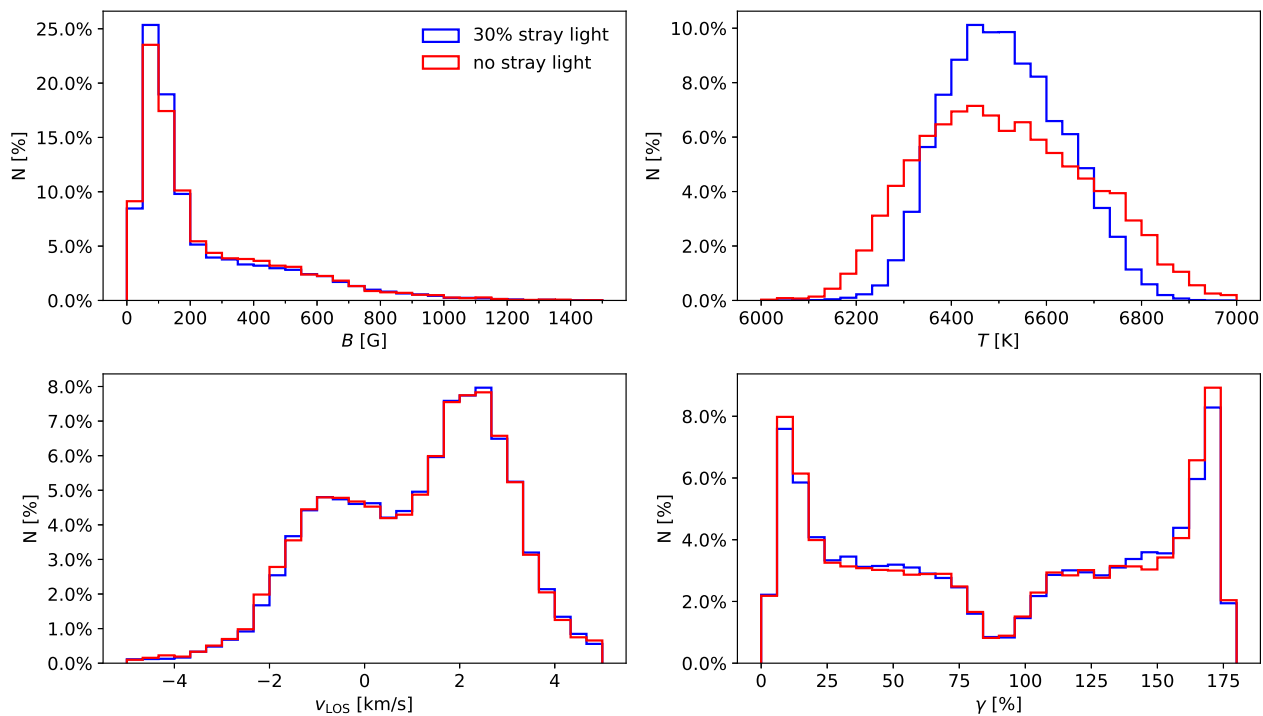}
    \caption{Impact of unpolarized stray light on the retrieval of the atmospheric parameters. Distributions of B, $T$, $v_{\mathrm{LOS}}$ and $\gamma$ returned from the inversions under S1, for V2 (\textit{red} lines) and V3 (\textit{blue} lines) degraded synthetic profiles (i.e. with and without stray light, respectively). $T$ is shown at $log\tau_{5000\mathrm{\AA}} = 0.5$.}
    \label{fig:degraded_stats_stray}
\end{figure*}

\begin{figure*}
    \centering
    \includegraphics[width=.9\textwidth]{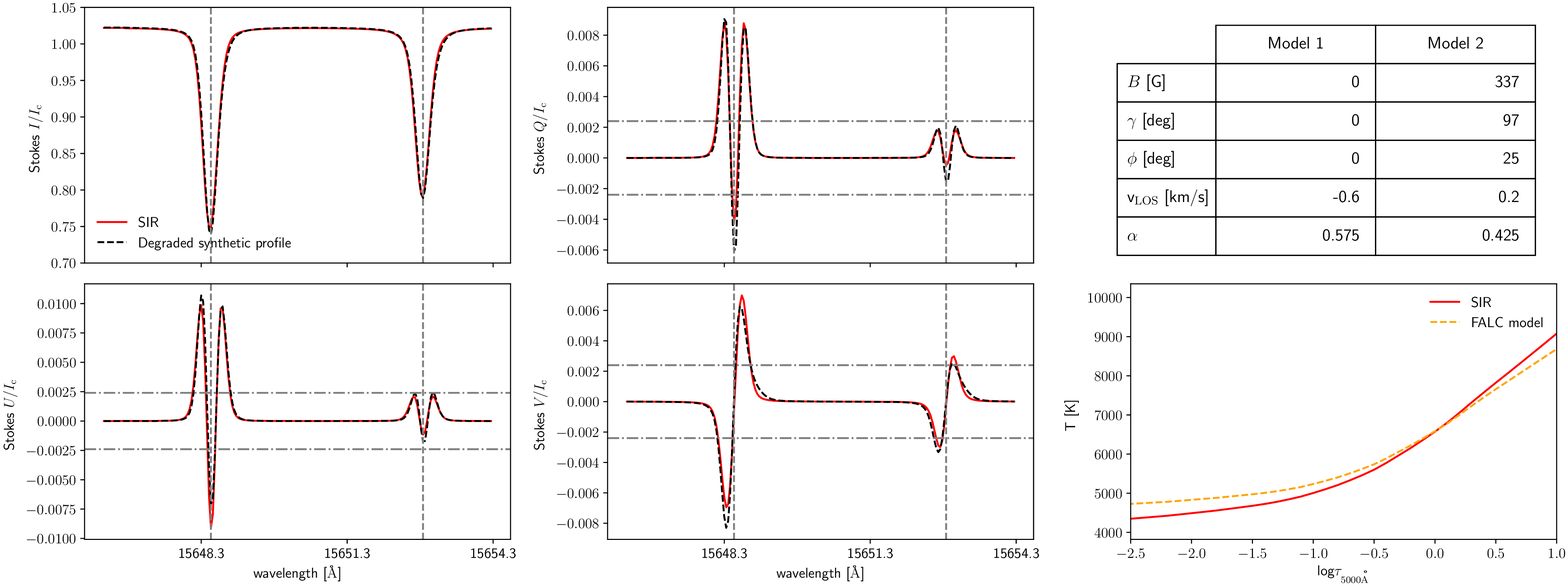}\vspace{0.1cm}
    \caption{Sample V3 profile, showing the full Stokes vector for the $15648.52/15652.87 \AA$ line pair in the left four panels, along with the S1 synthetic profiles derived from the SIR inversions. The horizontal (\textit{dot-dashed}) lines show the $3\sigma_n$ noise thresholds for the polarized Stokes vectors, while the vertical (\textit{dashed}) lines denote the rest wavelengths of each spectral line. On the right, the  retrieved atmospheric parameters from SIR is shown in the table, while the temperature as a function of optical depth is shown in the lower right plot with the original FALC input model. The arrow in Fig. \ref{fig:degraded_inclinations} indicates the spatial location of the sample profile.}
    \label{fig:Degraded_profile}
\end{figure*}

As the next step in our analysis, we inverted the degraded profiles again at the two noise levels but set any individual Stokes vector that did not have a maximum amplitude $>3\sigma_n$ before application of PCA to zero. Fig. \ref{fig:degraded_inclinations} shows the map of $\gamma$ inverted in this way alongside the noiseless case for V1 and V3. The maps at $\sigma_n = 8\times10^{-4}$ $I_{\mathrm{c}}$ look very similar to the regions of interest (ROIs) reported by \cite{Campbell} in terms of inclined fields flanked by opposite polarity vertical fields. The importance of adequate S/N is revealed by the map inverted with a noise level of $\sigma_n = 3\times10^{-4}$ $I_{\mathrm{c}}$, as our results suggest a much larger number of the inclined fields in particular should be measurable at this S/N compared to the map inverted with a noise level of $\sigma_n = 8\times10^{-4}$ $I_{\mathrm{c}}$. At the same time, by comparing maps at a given S/N between the figures, one can observe the importance of spatial resolution also; an increase in the spatial resolution of $0.23''$ enables a much larger number of inclined fields to be revealed, and the magnetic structures are revealed in increasing complexity.

While throughout this section of analysis we have been comparing V1 to V3 (as opposed to V1 to V2), and interpreting differences in terms of the difference in spatial resolution, it is also true that V3 additionally includes stray light as an additional variable. Fig. \ref{fig:degraded_stats_stray} shows the distributions of inverted atmospheric parameters for V2 and V3, where the only difference is the absence or inclusion of stray light, respectively. Evidently, the only atmospheric parameter impacted by the unpolarized stray light component is the $T$. The $T$ values for V2 have a wider spread. This is a predictable difference, as the inclusion of stray light reduces the continuum intensity contrast, which $T$ is very sensitive to. As \cite{milic2019} also found, $v_{\mathrm{LOS}}$ is invariant whether stray light is included or not. We additionally report that $B$ and $\gamma$ are also invariant.

Figure \ref{fig:Degraded_profile} shows a sample profile from the degraded cube (V3), whose location is marked by the arrow in Fig. \ref{fig:degraded_inclinations}. This profile is symmetric enough that scheme S1 is able to provide a very good fit to the full Stokes vector. The profile and the inferred depth-averaged atmospheric parameters resemble closely the vectors and values recorded and inferred in linear polarization features by \cite{Campbell}, in terms of being a highly inclined ($\gamma$ = $97^\circ$) weak field ($B$ = $337$ G) with a $\alpha B$ value of $143$ G. In principle, as both linear Stokes vectors are above the noise threshold, the azimuth should also be constrained (without resolution of the $180^\circ$ ambiguity). The same profile in V1 shows a more highly inclined ($\gamma$ = $100^\circ$) weaker field ($B$ = $300$ G) with a lower $\alpha B$ value of $106$ G. We cannot directly compare these values to MHD values. However, it is worth noting that there is a mix of opposite polarities in this region in the undegraded cube, showing that the spatial and spectral PSFs fundamentally alter how we observe and interpret the magnetic vector when we are not able to resolve the small-scale magnetic structures. Further, the amplitudes of each polarized Stokes vector is $2-3$ times smaller, highlighting the importance of adequate S/N in allowing us to constrain the magnetic vector in the IN at GREGOR resolutions, as well as the need for higher resolution telescopes.

\section{Discussion}

We have used a snapshot of a realistic MURaM simulation to show the extent to which we can accurately constrain information about the magnetic vector in the IN photosphere using inversion techniques. For inversions with constant parameters (scheme C, i.e. a Milne-Eddington atmosphere), we find the depth-averaged parameters are well constrained in congruence with previous studies (see, for instance, \cite{suarez2010}). The results with simple gradients (scheme G) in $B$, $\gamma$, and $v_{\mathrm{LOS}}$ show mixed results. The inclusion of a gradient in the $B$ and $v_{\mathrm{LOS}}$ seems to improve the $\chi^2$ and reduce the statistical error, suggesting that including gradients in $v_{\mathrm{LOS}}$ and $B$ may be warranted. However, the inclusion of gradients in the $\gamma$ in fact makes the statistical error slightly worse. The extent to which additional nodes can be added without the solutions becoming unconstrained or, even, oscillatory in optical depth is a question worthy of future investigation, as is the impact of the placement of nodes at specific optical depths \citep{Navarro2004,Navarro2008}. We find high correlations between the inverted and MHD parameters, and the depths at which $B$, $\gamma$, $v_{\mathrm{LOS}}$, and $T$ are highly correlated resembles the response functions very closely. It is clear that both $v_{\mathrm{LOS}}$ and $T$ are responsive to very deep layers of the atmosphere but we do not find evidence that $B$ is responsive to the same layers. The weakest correlation is observed in $\gamma$ and this parameter appears to be retrievable in the most narrow range of optical depths. Future studies that use SIR may be able to improve on these results if nodes can be placed at specifically selected optical depths, but ultimately more spectral lines will be required to probe the atmosphere at a greater range of optical depths. The recent publication of an expanded list of empirically determined atomic line parameters of spectral lines contained in the spectral region observed by GRIS \citep{lines2021} is worthy of further investigation to examine whether the inclusion of any of the spectral lines could improve the inversions. Otherwise, future multi-instrument telescopes (e.g. DKIST) will provide an opportunity to sample a more varied selection of spectral lines with sensitivities to a larger range of optical depths.

We then investigated how this model atmosphere would appear had it been observed at the spatial and spectral resolutions achievable by the GREGOR/GRIS-IFU and compared the results to observations reported by \cite{Campbell}. We found that the synthetic spectra closely resemble the real observations, with the caveat that a higher fraction of the FOV contains circular polarization signals when degraded than we observe in real observations. We studied a linear polarization feature (see Fig. \ref{fig:Degraded_profile}) which resembles those reported by \cite{Campbell} in terms of being flanked by opposite polarity vertical fields, resembling `loop-like' structures, but also in terms of its magnetic flux density ($\alpha B = 143$ G), which is very close to the values estimated in similar features observed by the GREGOR/GRIS-IFU. This shows that MURaM simulations are capable of showing close agreement with real observations, and the symbiosis between realistic MHD simulations and observations continues to prove essential. The highly significant impact of spatial and spectral degradation on the amplitudes of polarized Stokes profiles, and consequently our ability to measure them with modern polarimeters, is clear. Instrumental effects can result an atmosphere with nearly one fifth of pixels containing linear polarization signals high enough to be observed, to consequently have a reduced amplitude such that the signals are measurable in less than a few percent of the FOV, depending on the S/N and tolerance for noise. At the same time, we found that $B$, $\gamma$, and $v_{\mathrm{LOS}}$ are invariant, and that the contrast in $T$ is reduced, when an unpolarized stray light contribution is included.

As the linear polarization signals are typically so close to the noise level at GREGOR resolutions, the inclination and azimuth values returned from inversions are highly susceptible to noise contamination in the IN photosphere. As \cite{Campbell} found, this makes it very difficult to observe the temporal evolution of weak and inclined fields; their linear polarization signals can vanish between frames. The authors elected not only to restrict analysis to pixels which had just one polarization signal greater than a noise threshold, but also to set any individual Stokes vector with a maximum amplitude lower than this noise threshold to zero (as in Fig. \ref{fig:degraded_inclinations}). This was intended to restrict analysis to only those Stokes vectors that were actually measured, and is necessary as inversion codes like SIR will insert spurious Stokes vectors into pure photon noise. Consider the case where we have measured only Stokes $V$ above a noise threshold. If we have not confidently measured at least Stokes $Q$ or $U$, the only information we would have about the linear polarization signals is that their amplitude is not larger than the noise threshold. However, the amplitude of the Stokes $Q$ or $U$ signals could be any value between the noise threshold and zero. Depending on the relative amplitude  of the corresponding Stokes $V$ profile, this can result in large errors in $\gamma$ (as in Fig. \ref{fig:degraded_stats}). This difficulty is even further encountered when trying to measure azimuths accurately, as one would need to measure both linear Stokes vectors confidently. Faced with this problem, we are left with a number of unpalatable options. We could:
\begin{enumerate}
    \item Attempt to bin pixels spatially, spectrally, or temporally to reduce the noise level, as in \cite{lagg2016,Campbell}, at the risk of fundamentally altering the atmospheres due to Zeeman mixing of Stokes signals,
    \item Set the Stokes vectors that are not measured above an appropriate confidence level to zero, and accept that we can only determine whether fields are predominantly vertical or horizontal in most pixels, or
    \item Invert the profiles with noise, knowing that the inversion will produce an inflated number of transverse fields beyond what has actually been measured in the observations, that the retrieved magnetic parameters may have large errors induced by noise, and thus any temporal variation may be spurious.
\end{enumerate}
If inversions are conducted with both the latter approaches, we can apply upper and lower limits, respectively, on the number of transverse fields. The errors introduced by noise, if similar in magnitude to those in Fig. \ref{fig:degraded_stats} and Table \ref{table:stds}, might be acceptable for retrieving statistical results. However, when trying to constrain the magnetic vector in small-scale features by following the dynamics in given pixels or groups of pixels between frames the errors could be much larger than the medians we report here. Therefore, it naturally follows that perhaps future observers should target a higher S/N. Future observers attempting to observe these magnetic fields with larger telescopes will also have to consider the implications of these results. Hopefully, with higher spatial resolution the measured polarization signals will have a higher amplitude, as there will be less mixing of opposite polarities. For this reason, next generation solar telescopes like DKIST and the European Solar Telescope \citep{EST} will revolutionize our understanding of IN magnetic fields. 



\begin{acknowledgements}
     We are grateful to have benefited from the insight and expertise of the anonymous referee who helped improve this manuscript. R. J. C. additionally thanks all speakers, organizers and participants of the SOLARNET-funded Summer School held in September 2019 in the Università della Svizzera italiana, Lugano, Switzerland. In particular, Dr. Juan Manuel Borrero and Dr. Adur Pastor Yabar are thanked for useful discussions that helped guide the approach taken in this study. R. J. C. thanks Dr. Robert Ryans for IT support and assistance in utilizing QUB's high performance computing (HPC) facilities. Dr. Ricardo Gafeira is thanked for helping us make use of his parallelized SIR wrapper, which is highly recommended. R. J. C. acknowledges support from the Northern Ireland Department for the Economy (DfE) for the award of a PhD studentship. This research has received financial support from the European Union’s Horizon $2020$ research and innovation program under grant agreement No. $824135$ (SOLARNET). M. M. acknowledges support from the Science and Technology Facilities Council (STFC) under grant No. ST$/$P000304$/$1 $\&$ ST$/$T00021X$/$1. The $1.5$-meter GREGOR solar telescope was built by a German consortium under the leadership of the Leibniz-Institute for Solar Physics (KIS) in Freiburg with the Leibniz Institute for Astrophysics Potsdam, the Institute for Astrophysics Göttingen, and the Max Planck Institute for Solar System Research in Göttingen as partners, and with contributions by the Instituto de Astrof\'isica de Canarias and the Astronomical Institute of the Academy of Sciences of the Czech Republic.
\end{acknowledgements}

%
%

\bibliographystyle{aa.bst} 
\bibliography{bib} 

\end{document}